\documentclass[preprintnumbers, floatfix, preprintnumbers, letterpaper, superscriptaddress,nofootinbib]{revtex4}
\pdfoutput=1
\usepackage{graphicx}
\usepackage{microtype}
\usepackage{amsmath}
\usepackage{amssymb}
\usepackage{subfigure}
\usepackage{hyperref}
\usepackage{url}
\usepackage{xcolor}
\usepackage{color}
\usepackage{mathrsfs}
\usepackage{calrsfs}
\usepackage{amsfonts}
\usepackage{latexsym}
\usepackage{ragged2e}
\usepackage{epstopdf}
\usepackage{textcomp}
\usepackage{phaistos}
\usepackage{ulem}

\makeatletter
\renewcommand\@makefnmark{\hbox{\@textsuperscript{\normalfont\color{purple}\@thefnmark}}}
\renewcommand\@makefntext[1]{%
  \parindent 1em\noindent
            \hb@xt@1.8em{%
                \hss\@textsuperscript{\normalfont\@thefnmark}}#1}
\makeatother

\usepackage{caption}
\DeclareCaptionJustification{justified}{\leftskip=0pt \rightskip=0pt \parfillskip=0pt plus 1fil}
\definecolor{vividviolet}{rgb}{0.62, 0.0, 1.0}
\definecolor{amaranth}{rgb}{0.9, 0.17, 0.31}
\definecolor{palatinateblue}{rgb}{0.15, 0.23, 0.89}
\definecolor{brightpink}{rgb}{1.0, 0.0, 0.5}
\definecolor{cornflowerblue}{rgb}{0.39, 0.58, 0.93}
\definecolor{deepcarminepink}{rgb}{0.94, 0.19, 0.22}
\definecolor{radicalred}{rgb}{1.0, 0.21, 0.37}

\hypersetup{ linktoc=all,
    colorlinks, linkcolor={palatinateblue},
    citecolor={brightpink}, urlcolor={amaranth}
}

\graphicspath{{Images/}}

\def\sideremark#1{\ifvmode\leavevmode\fi\vadjust{\vbox to0pt{\vss
 \hbox to 0pt{\hskip\hsize\hskip1em
 \vbox{\hsize1.5cm\tiny\raggedright\pretolerance10000
 \noindent #1\hfill}\hss}\vbox to8pt{\vfil}\vss}}}%
\begin{document}

\title{Scalar Hairy Black Holes with Inverted Mexican Hat Potential}

\author{Xiao Yan \surname{Chew}}
\email{xychew998@gmail.com}
\affiliation{Department of Physics, School of Science, Jiangsu University of Science and Technology, 212100, Zhenjiang, Jiangsu Province, People's Republic of China }

\author{Kok-Geng \surname{Lim}}
\email{K.G.Lim@soton.ac.uk}
\affiliation{Smart Manufacturing and Systems Research Group, University of Southampton Malaysia, 79100 Iskandar Puteri, Malaysia}

\begin{abstract}
We numerically construct the asymptotically flat solutions of hairy black holes supported by a symmetric inverted Mexican hat potential with a local minimum and two degenerate global maxima of a real scalar field that contains a quartic self-interaction term. The solutions of hairy black holes emerge from the Schwarzschild black hole when the non-trivial scalar field exists outside the event horizon. Therefore, we perform a comprehensive study on the properties of the hairy black holes such as the area of horizon, the Hawking temperature, the innermost stable circular orbit, the photon sphere, etc. We also numerically study their linear stability in the mode analysis, hence finding that they are unstable against the linear perturbation. 
\end{abstract}

\maketitle

\section{Introduction}

In General Relativity (GR) the electrovacuum black holes only possess three global charges which are the mass, electrical charge and angular momentum. Hence, they satisfy the no-hair theorem which states that black holes can only be described by the corresponding three global charges \cite{Bekenstein:1995un,Israel:1967wq,Ruffini:1971bza}. However, when the black holes are supported by the matter fields outside their event horizon, they are known as hairy black holes, which can be constructed with the circumvention of no-hair theorem. The term ``hair" is coined by J. Wheeler and referred to the extra global charges associated with the matter fields that can be possibly possessed by the hairy black holes. 

In GR, the hairy black holes can exhibit some new behaviours which deviate from the electrovacuum black holes in the strong gravity regime, for example when a hairy black hole is supported by a real scalar field, it can possess a nontrivial scalar hair outside the event horizon by undergoing a mechanism which is called the ``spontaneous scalarization" (SS). The concept of SS was first proposed by T. Damour and G. Esposito-Far\`ese \cite{Damour:1993hw} to predict the deviation of properties for the neutron stars from GR in the framework of Scalar-Tensor theory which non-minimally couples a scalar function with the Ricci scalar but becomes non-distinguishable in the weak gravity regime. The SS of hairy black holes can be realized by non-minimally coupling various forms of the scalar function $f(\phi)$ with the Maxwell field in the Einstein-Maxwell-scalar (EMs) theory \cite{Gibbons:1987ps,Garfinkle:1990qj,Dobiasch:1981vh,Gibbons:1985ac,Kallosh:1992ii,Herdeiro:2018wub,Fernandes:2019rez,Myung:2018vug,Myung:2018jvi,Myung:2019oua,Astefanesei:2019pfq,Blazquez-Salcedo:2019nwd,Astefanesei:2019qsg,Myung:2020dqt,Astefanesei:2020xvn,LuisBlazquez-Salcedo:2020rqp,Blazquez-Salcedo:2020nhs} and with the geometrical terms such as the Gauss-Bonnet term in the Einstein-Gauss-Bonnet-scalar (EGBs) theory \cite{Mignemi:1992nt,Kanti:1995vq,Torii:1996yi,Pani:2009wy,Ayzenberg:2014aka,Maselli:2015tta,Kleihaus:2011tg,Kleihaus:2015aje,Sotiriou:2014pfa,Doneva:2017bvd,Silva:2017uqg,Antoniou:2017acq,Antoniou:2017hxj,Herdeiro:2018wub,Andreou:2019ikc,Cunha:2019dwb,Collodel:2019kkx,Dima:2020yac,Herdeiro:2020wei,Berti:2020kgk}. 

Nevertheless, one may overlook the fact that there could be the simplest and most direct method to construct the scalar hairy black holes in GR, that is to direct minimally couple the Einstein gravity with a potential $V(\phi)$ of a scalar field. By carefully introducing the form of $V(\phi)$ which is associated with the profile of $V(\phi)$, it is possible to circumvent the no-hair theorem and regularize the scalar field at the horizon, such that the solutions of hairy black holes can emerge from Schwarzschild black hole when the scalar field exists at the horizon, for example, Refs. \cite{Corichi:2005pa,Chew:2022enh} have employed an asymmetrical shape of scalar potential to construct the hairy black holes where it contains a local maximum, a local minimum  and a global minimum to describe the quantum tunnelling process from the false vacuum (local minimum) to the true vacuum (global minimum) in the cosmology \cite{Coleman:1980aw}. In addition, Ref. \cite{Gubser:2005ih} has constructed a hairy black hole using a symmetrical shape of scalar potential with a local minimum at zero and two degenerate global maxima which looks like an inverted Mexican hat and contains a quartic self-interaction term. However, we find that \cite{Gubser:2005ih} didn't study the properties of black holes systematically and the corresponding results haven't been published yet to the best of our knowledge, hence it would be interesting that we perform a comprehensive study on the properties of hairy black holes and report them in this paper. Other construction of scalar hairy black holes can be found in Refs. \cite{Nucamendi:1995ex,Bechmann:1995sa,Dennhardt:1996cz,Bronnikov:2001ah,Martinez:2004nb,Winstanley:2005fu,Nikonov:2008zz,Anabalon:2012ih,Gao:2021ubl,Karakasis:2021rpn,Karakasis:2022xzm,Karakasis:2023ljt,Lan:2023cvz,Cadoni:2015gfa,Radu:2011uj}.

Furthermore, the recent two major astrophysical events have brought us the exciting prospects in the black hole physics, where we can search for the existence of hairy black holes in the future from these astrophysical signatures. The two major astrophysical events are the detection of the emission of gravitational waves from the merger of binary black holes by the LIGO-VIRGO-KAGRA (LVK) collaboration \cite{LIGOScientific:2016aoc,LIGOScientific:2017vwq,LIGOScientific:2017ync,LIGOScientific:2019fpa,Barack:2018yly,LIGOScientific:2020ibl} and the image of supermassive black holes in the center of M87 \cite{EventHorizonTelescope:2019dse,EventHorizonTelescope:2019ths,EventHorizonTelescope:2019pgp,EventHorizonTelescope:2019ggy} and Sgr $\text{A}^{*}$ \cite{EventHorizonTelescope:2022wkp,EventHorizonTelescope:2022xqj,EventHorizonTelescope:2022urf,EventHorizonTelescope:2022exc} captured by the Event Horizon Telescope (EHT). Hence, our motivation is to construct the solutions of hairy black holes as a first step in order to study their astrophysical signatures.

This paper is organized as follows. In Sec.~\ref{sec:th}, we briefly introduce our theoretical setup comprising the Lagrangian and the metric ansatz in Sec.~\ref{sec:thz}. Then, we briefly justify the existence of hairy black holes with a few simple analyses in Sec.~\ref{ssec:remarks}; derive the set of coupled differential equations and study the asymptotic behaviour of the functions in Sec.~\ref{ssec:ode}. In Sec.~\ref{sec:prop}, we briefly introduce the properties of hairy black holes, which include the area of horizon, the Hawking temperature, Ricci scalar and Kretschmann scalar in Sec.~\ref{ssec:bprop}. We also study the geodesics of test particles in the vicinity of hairy black holes in Sec.~\ref{ssec:geo} and the linear stability of the hairy black holes in the mode analysis in Sec.~\ref{ssec:lin}. We present and discuss our numerical results in Sec.~\ref{sec:res}. Finally, we conclude our research work in Sec.~\ref{sec:con}.

\section{Theoretical Setting} \label{sec:th}

\subsection{Theory and Ans\"atze} \label{sec:thz}

\begin{figure}
\centering
\mbox{
(a)
 \includegraphics[trim=50mm 170mm 20mm 20mm,scale=0.58]{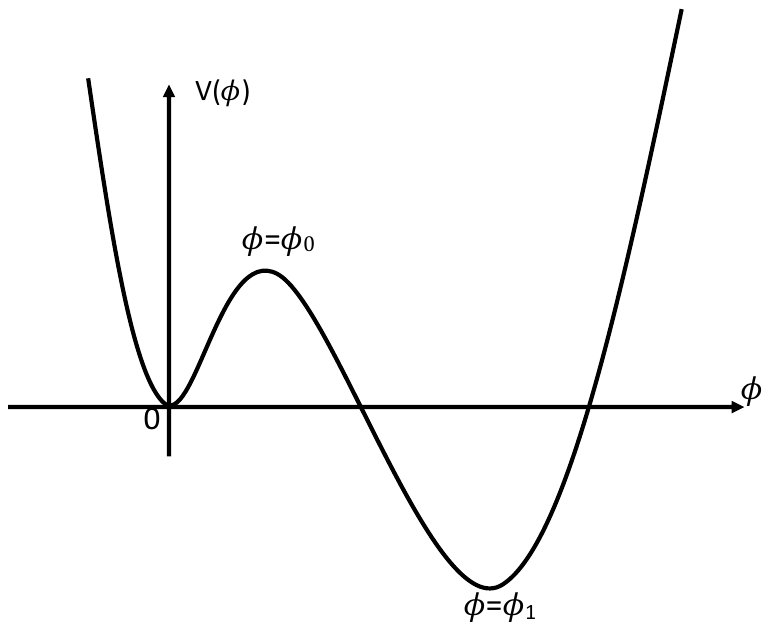}
(b)
 \includegraphics[angle =0,scale=0.51]{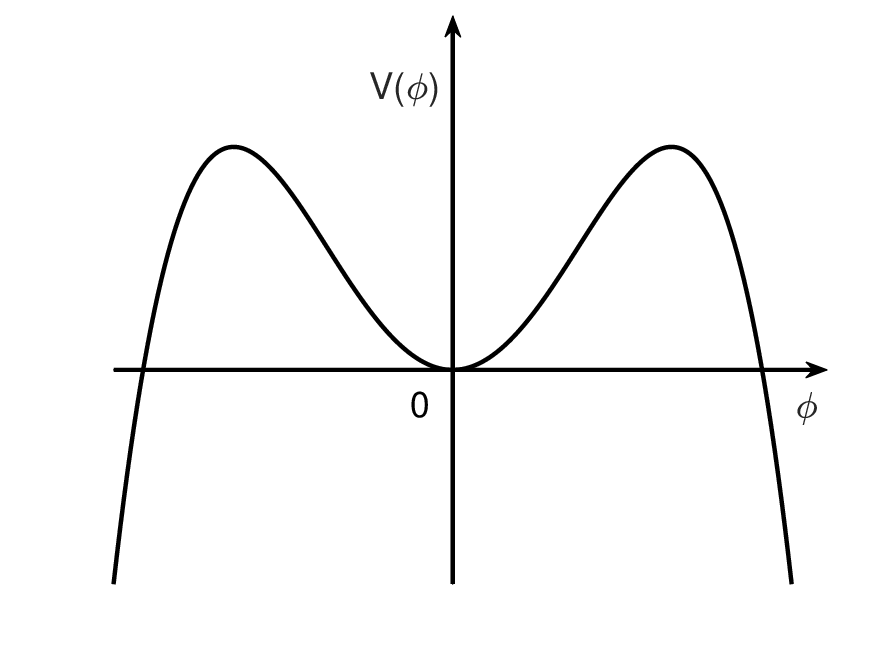}
}
\caption{(a) The authors have considered the asymmetric shape of scalar potential $V(\phi)$ with a local minimum at $\phi=0$, a local maximum at $\phi=\phi_0$ and a global minimum at $\phi=\phi_1$ to construct the hairy black holes \cite{Chew:2022enh}. (b). Ref. \cite{Gubser:2005ih} has considered the symmetric potential which looks like the Mexican hat with two degenerate global maxima at $\phi_{\text{max}}=\pm \sqrt{\mu/\Lambda}$ to construct hairy black holes.}
\label{plot_Vphi}
\end{figure}

We consider the Einstein-Klein-Gordon (EKG) system where a scalar potential $V(\phi)$ of a scalar field $\phi$ minimally couples with the Einstein gravity,  
\begin{equation} \label{EHaction}
 S=  \int d^4 x \sqrt{-g}  \left[  \frac{R}{16 \pi G} - \frac{1}{2} \nabla_\mu \phi \nabla^\mu \phi - V(\phi) \right]  \,,
\end{equation}
and the form of scalar potential $V(\phi)$ is given explicitly as a sum of polynomial for $\phi$,
\begin{equation}
  V(\phi)  = \sum_{i=0}^{\infty} k_i \phi^i \,,
\end{equation}
with $k_i$ as an arbitrary real-valued constant. Note that Ref. \cite{Gao:2021ubl} has considered the above scalar potential $V(\phi)$ to construct the hairy black holes but didn't study their properties systematically.  If we truncate $V(\phi)$ up to $i=4$, then it reads
\begin{equation}
  V(\phi)  =  k_4 \phi^4 + k_3 \phi^3 + k_2 \phi^2 + k_1 \phi \,,
\end{equation}
with $k_0=0$ such that the scalar field vanishes at the spatial infinity. If $k_3 \neq 0$, the shape of this type of $V(\phi)$ looks asymmetric as shown in Fig.  \ref{plot_Vphi} (a) which has been considered by the authors in \cite{Chew:2022enh} with a specific form of parametrization for $k_i$ to construct the solutions of hairy black holes. If the coefficients of odd power terms vanish, i.e., $k_1=k_3=0$, then the shape of $V(\phi)$ is symmetric and looks like an inverted Mexican hat with a local minimum at $\phi=0$ and two degenerate global maxima at $\phi_{\text{max}}=\pm \sqrt{\mu/\Lambda}$ as shown in Fig. \ref{plot_Vphi} (b), where its explicit form is given by
\begin{equation}
 V(\phi) = -\Lambda \phi^4 + \mu \phi^2 \,, \label{ppot}
\end{equation}
with $\Lambda$, $\mu$ are the constants. This $V(\phi)$ has been considered in Ref. \cite{Gubser:2005ih} to construct the hairy black holes but their properties haven't been studied systematically for a few decades ago. However, we couldn't find that the corresponding results have been published yet in any other journals. Therefore, in this paper, we also construct such hairy black holes but perform a comprehensive study to analyze their properties.  
   
We then obtain the Einstein equation and Klein-Gordon (KG) equation from the variation of the action with respect to the metric and scalar field, respectively
\begin{align} 
 R_{\mu \nu} - \frac{1}{2} g_{\mu \nu} R &= 8 \pi G \left(  -   \frac{1}{2} g_{\mu \nu} \nabla_\alpha \phi \nabla^\alpha \phi - g_{\mu \nu} V + \nabla_\mu \phi \nabla_\nu \phi    \right) \,, \label{EFE} \\ 
 \nabla_\mu \nabla^\mu \phi  &=  \frac{d V}{d \phi} \,. \label{KGeqn}  
\end{align}

We employ the following metric with spherically symmetric property as the Ansatz to construct the hairy black holes,
\begin{equation}  \label{line_element}
ds^2 = - N(r) e^{-2 \sigma(r)} dt^2 + \frac{dr^2}{N(r)} + r^2  \left( d \theta^2+\sin^2 \theta d\varphi^2 \right) \,, 
\end{equation}
where $N(r)=1-2 m(r)/r$ with $m(r)$ is the Misner-Sharp mass function \cite{Misner:1964je}. We can read off the ADM mass of the hairy black holes with the condition $m(\infty)=M$.

\subsection{Remarks on the Existence of Hairy Black Holes} \label{ssec:remarks}

Here we briefly discuss the existence of hairy black holes. The trivial solution for the EKG system is the Schwarzschild black hole when the scalar field $\phi$ does not exist. When the scalar field exists but the scalar potential vanishes $(V=0)$, the solution for the metric is also the Schwarzschild black hole but the scalar field diverges at the horizon with the form 
\begin{equation}
   \phi'(r) \sim \frac{1}{1-\frac{2 m}{r}}  \Rightarrow \phi(r) \sim \ln \left( \frac{2M}{r} -1 \right) \,.
\end{equation}
Thus, one can carefully introduce $V(\phi)$ to regularize the scalar field at the horizon such that the hairy black holes are regular everywhere in the spacetime. Moreover, we perform a few simple analyses to justify the existence of hairy black holes by merely referring to the KG equation without using the Einstein equation \cite{Herdeiro:2015waa}. Firstly, we multiply $\phi$ to both sides of the KG equation (Eq. \eqref{KGeqn}) and integrate it by parts from the black hole horizon to infinity, since the boundary term vanishes at the horizon and infinity, hence we are left with the following integral,
\begin{equation}
 \int_{r_H}^{\infty} d^4 x \sqrt{-g} \left[  \nabla_\mu \phi \nabla^\mu \phi + \phi  \frac{d V}{d \phi}   \right] =0 \,.
\end{equation}

The term $ \nabla_\mu \phi \nabla^\mu \phi = \left( \partial_r \phi \right)^2  \geq 0$ because $\phi$ is spherically symmetric and stationary which is independent of time. Thus, the integral implies the condition $\phi \frac{d V}{d \phi} \leq 0$ has to be satisfied. Here, we assume $\phi$ is always greater than zero and nodeless in our case, this implies that $\frac{d V}{d\phi}<0$, which can be satisfied since $\frac{d V}{d\phi}= -4 \Lambda \phi^3 + 2 \mu \phi < 0$. 

Secondly, we multiply $\frac{d V}{d \phi}$ to both sides of the Eq. \eqref{KGeqn} and repeat again the above procedure, obtaining
\begin{equation}
 \int_{r_H}^{\infty} d^4 x \sqrt{-g} \left[  \frac{d^2 V}{d \phi^2} \nabla_\mu \phi \nabla^\mu \phi  + \left(   \frac{d V}{d \phi}  \right)^2    \right] = 0   \,.
\end{equation}
The above integral obviously implies that the condition $\frac{d^2 V}{d \phi^2} < 0 $ has to be satisfied in order to make the integral vanishes non-trivially, this indicates that the profile of $V$ has to be concave-down and possess possibly at least a local maximum. In our case, the condition $\frac{d^2 V}{d \phi^2} = -12 \Lambda \phi^2 + 2 \mu  < 0 $ can also be satisfied. 

Furthermore, we can inspect the weak energy condition since it can be possibly violated with $V < 0$ in some regions of $\phi$, 
\begin{equation} \label{WEC}
 \rho = - T^t_{t} = \frac{N}{2} \phi'^2 + V  \,.
\end{equation}
The violation of the weak energy condition could lead to the violation of the strong energy condition but the opposite may not necessarily true.

\subsection{Ordinary Differential Equations (ODEs)} \label{ssec:ode}

The substitution of Eq. \eqref{line_element} into the Eqs. \eqref{EFE} and \eqref{KGeqn} yields a set of second-order and nonlinear ODEs for the functions,
\begin{equation}
m' = 2 \pi G \left( N \phi'^2 + 2 V \right) \,, \quad \sigma' = - 4 \pi G r \phi'^2 \,,    \quad
\left(  e^{- \sigma} r^2 N \phi' \right)' = e^{- \sigma} r^2  \frac{d V}{d \phi} \,,
\end{equation}
where the prime denotes the derivative of the functions with respect to the radial coordinate $r$. Here we only consider the solutions of hairy black holes for the range of $r$ from the horizon radius $r_H$ to infinity. Although the form of ODEs looks very simple, it is almost impossible to obtain the analytical solutions for the hairy black holes, hence we integrate the ODEs numerically from the horizon to the infinity.

In order to construct the hairy black hole solutions with globally regular, we demand all the functions and their derivatives have to be finite at the horizon $r_H$. With this regularity condition, we can extract the asymptotic behaviour of the functions at the horizon and the infinity in the form of series expansion. At the horizon, the corresponding series expansion with few leading terms are given by
\begin{align}
 m(r) &= \frac{r_H}{2}+ m_1 (r-r_H) + O\left( (r-r_H)^2 \right) \,, \label{m_ex} \\
\sigma(r) &= \sigma_H + \sigma_1   (r-r_H) + O\left( (r-r_H)^2 \right)  \,, \\
 \phi(r) &= \phi_H +  \phi_{H,1}  (r-r_H) + O\left( (r-r_H)^2 \right)  \,,
\end{align} 
where
\begin{equation}
   m_1 = 4 \pi G r^2_H  V(\phi_H)  \,, \quad  \sigma_1 = -  4 \pi G r_H \phi_{H,1} \,, \quad   \phi_{H,1}= \frac{r_H \frac{d V(\phi_H)}{d \phi}}{1-8 \pi G r_H^2 V(\phi_H)}  \,.
\end{equation} 
Here $\sigma_H$ and $\phi_H$ are the values of $\sigma$ and $\phi$ at the horizon, respectively. While at the infinity $(r\to \infty)$, we impose the condition of asymptotical flatness by requiring the scalar field to vanish, then the leading terms for the corresponding series expansion are given by the following expressions
\begin{align}
    m(r) &= M  + \tilde{m}_1 \frac{\exp{(- 2 m_{\text{eff}} r)}}{r} + ...\, , \\
    \sigma(r) &= \tilde{\sigma}_1 \frac{ \exp{(- 2 m_{\text{eff}} r)}}{r}  +   ... \, , \\
    \phi(r) &= \tilde{\phi}_{H,1}  \frac{ \exp{(- m_{\text{eff}} r)} }{r} + ... \, ,
\end{align}
where $\tilde{m}_1$, $\tilde{\sigma}_1$ and $\tilde{\phi}_{H,1}$ are constants, $M$ is the total mass of the configuration. 
Note that the effective mass of the scalar field is given by $m_{\text{eff}}=\mu$. 

Moreover, the ODEs are solved by the professional ODE solver package Colsys \cite{Ascher:1979iha} and Matlab package bvp4c \cite{Shampine:2001}. Colsys employs the Newton-Raphson method to solve the boundary value problem for a set of nonlinear ODEs by providing the adaptive mesh refinement to generate the solutions to have more than 1000 points with high accuracy and the estimation of errors of solutions. Whereas bvp4c is an adaptive mesh BVP solver that implements the three-stage Lobatto IIIa collocation formula. 

\section{Properties of the Scalar Hairy Black Holes} \label{sec:prop}

In this section, we briefly introduce several properties of hairy black holes that will be studied in this paper.

\subsection{Basic Properties}   \label{ssec:bprop}

We can study the behaviour of horizon for the hairy black holes with the two basic quantities which are the area of horizon $A_H$ and the Hawking temperature $T_H$,
\begin{equation}
A_H = 4 \pi r^2_H \,, \quad T_H = \frac{1}{4 \pi} N'(r_H) e^{-\sigma_H}   \,.
\end{equation}
In order to compare our hairy black holes with a known solution of electrovacuum black holes, which is the Schwarzschild black hole in this case, then we rescale $A_H$ and $T_H$ in the following way,
\begin{equation}
 a_H = \frac{A_H}{16 \pi M^2} \,, \quad t_H = 8 \pi T_H M \,.
\end{equation}
Hence both values $a_H$ and $t_H$ are equal to unity when the black hole is a Schwarzschild black hole.

We can also inspect the Ricci scalar $R$ and Kretschmann scalar $K=R_{\alpha \beta \gamma \delta} R^{\alpha \beta \gamma \delta}$ of the hairy black holes, they can be derived explicitly as
\begin{align}
 R &= -N'' + \frac{3 r \sigma'-4}{r} N' + \frac{2 \left( 2 r N \sigma' - N +1 + r^2 N\sigma'' - r^2 N \sigma'^2  \right)}{r^2} \,, \\
 K &= \left(  3 \sigma' N' + 2 N \sigma'' - N'' -2 N \sigma'^2  \right)^2    +  \frac{2}{r^2} \left( N'-2 N \sigma'  \right)^2 + \frac{2 N'^2}{r^2} + \frac{4 (N-1)^2}{r^4}  \,.
\end{align}
Thus, we find that $R$ and $K$ are finite with few leading orders using series expansion of the functions at the horizon, 
\begin{align}
  R &=  -\frac{2 m_1 \left( 3 r_H \sigma_1 -2 \right)}{r^2_H} + \frac{3 \sigma_1 + 4 m_2}{r_H} +  O\left( (r-r_H) \right)  \,, \\
  K &=  \frac{16 m^2_2}{r^2_H} - \frac{8 \left(  -2 + 6 m_1 \sigma_1 r_H + 4 m_1 - 3 \sigma_1 r_H  \right)}{r^3_H}   \nonumber \\
& \quad + \frac{12-32 m_1 + 48 m^2_1 \sigma_1 r_H + 36 m^2_1 \sigma^2_1 r^2_H + 32 m^2_1 + 9 \sigma^2_1 r^2_H + 12 \sigma_1 r_H - 36 \sigma^2_1 r^2_H m_1 - 48 m_1 \sigma_1 r_H}{r^4_H}   +  O\left( (r-r_H) \right)  \,,
\end{align}
where $m_2$ is the coefficient of second-order term in Eq. \eqref{m_ex}. The Ricci scalar of Schwarzschild black hole vanishes but its Kretschmann scalar is given by $K=12r_H^2/r^6$. 

\subsection{Geodesics of Test Particles Around the Hairy Black Holes} \label{ssec:geo}

In this subsection, we study the motion of test particles in the vicinity of hairy black holes, since it carries some important information about how the hairy black hole could distort the spacetime in the strong gravity regime that affects the motion of test particles which is useful for the investigation of image of black hole shadow in the future. Hence, we begin with the Lagrangian $\mathcal{L}$ is given by
\begin{equation}
 \mathcal{L} = \frac{1}{2} \dot{x}^\mu \dot{x}_\mu = \epsilon \,,
\end{equation} 
where $\epsilon=0$ refers to massless particle and $\epsilon=-1$ refers to massive particle. The dot denotes the derivative of a function with respect to an affine parameter.

Since the spacetime is static and stationary, it possesses two conserved quantities which are the energy $E$ and angular momentum $L$, 
\begin{equation}
 E = - \frac{\partial E}{\partial \dot{t}} = e^{-2 \sigma} N \dot{t} \,, \quad  L =  \frac{\partial E}{\partial \dot{\varphi}} = r^2 \dot{\varphi} \,. 
\end{equation}
In particular, we concentrate on the motion of test particles on the equatorial plane $(\theta=\pi/2)$, then the radial equation $\dot{r}$ reads
\begin{equation}
 e^{-2 \sigma} \dot{r}^2 =    E^2 -  V_{\text{eff}}(r)  \,, 
\end{equation}
where the effective potential $V_{\text{eff}}(r)$ is given by
\begin{equation}
 V_{\text{eff}}(r) = e^{-2 \sigma} N \left(  \frac{L^2}{r^2} - \epsilon \right) \,.
\end{equation}

Therefore, we can study the innermost stable circular orbit (ISCO) of a massive test particle where it can still move in the stable circular orbit at a minimal radius in the bulk of hairy black hole. The ISCO can be determined by the following condition,
\begin{equation}
  \frac{d V_{\text{eff}} }{d r} =   \frac{d^2 V_{\text{eff}} }{d r^2} = 0 \,. 
\end{equation}
The explicit forms for the above expressions are given by
\begin{align}
  \frac{d V_{\text{eff}} }{d r} &=  \left( \frac{N'}{N} - 2 \sigma  \right)  V_{\text{eff}} - \frac{2 L^2 e^{-2 \sigma } N}{r^3}    \,, \\
 \frac{d^2 V_{\text{eff}} }{d r^2} &= \left(  \frac{N''}{N} - \frac{4 \sigma' N'}{N} + 4 N \sigma'^2 - 2 \sigma''  \right)    V_{\text{eff}} + 2 L^2 e^{-2\sigma} \left( \frac{3 N}{r^4} + \frac{4 N \sigma'}{r^3} - \frac{2 N'}{r^3}  \right) \,.
\end{align}
Hence, setting $\frac{d V_{\text{eff}} }{d r} =   \frac{d^2 V_{\text{eff}} }{d r^2} = 0$ yields 
\begin{equation}
\frac{N''}{N} - \frac{2 N'^2}{N^2} - 2 \sigma'' + \left(4 \sigma' + \frac{3}{r }    \right) \frac{N'}{N} - 2 \left( 2 \sigma' + \frac{3}{r}   \right) \sigma'    = 0 \,. 
\end{equation}
The location of ISCO for a massive test particle can be obtained by solving the above equation numerically. Moreover, the location of the photon sphere can be obtained numerically by solving $\frac{d V_{\text{eff}} }{d r} = 0$ with $\epsilon=0$.

\subsection{The Radial Perturbation} \label{ssec:lin}

Here we investigate the linear stability of hairy black holes, we begin our investigation by perturbing the background metric and scalar field, respectively as 
\begin{eqnarray}
 ds^2 &=& - N(r) e^{-2 \sigma(r)} \left[  1 + \epsilon e^{-i \omega t} F_t(r)  \right] dt^2 + \frac{1}{N(r)} \left[ 1+ \epsilon e^{-i \omega t} F_r(r)   \right] dr^2 + r^2 \left( d \theta^2+\sin^2 \theta d\varphi^2 \right) \,, \\
 \Phi &=& \phi(r) + \epsilon  \Phi_1 (r) e^{-i \omega t} \,,
\end{eqnarray}
where $F_t(r)$,  $F_r(r)$ and $\Phi_1 (r)$ are the small perturbations to the background solutions. The substitution of the above Ansatz to the Einstein equation and KG equation yields a set of ODEs with the first order in $\epsilon$, 
\begin{eqnarray}
 F_r &=& 8 \pi G r \Phi_1 \phi'   \,, \label{ODEper1} \\
 F'_t &=& - F'_r + 16 \pi G r \Phi'_1 \phi'   \,, \label{ODEper2} \\
\Phi''_1 &=& \left(  \sigma' - \frac{N'}{N} -\frac{2}{r} \right)  \Phi'_1 + \left(  \frac{1}{N}  \frac{d^2 V}{d \phi^2}  - \omega^2 \frac{e^{2 \sigma}}{N^2}  \right)  \Phi_1 + \frac{F_r}{N}\frac{d V}{d \phi} + \frac{ F'_r - F'_t }{2} \phi'  \,. \label{ODEper3}
\end{eqnarray}
It is obvious that only Eq. \eqref{ODEper3} is independent since Eqs. \eqref{ODEper1} and \eqref{ODEper2} are dependent. Next, we transform $\Phi''_1 $ into a master equation which is Schr\"odinger-like by defining $Z(r) = r \Phi_1(r)$,
\begin{equation}
 \frac{d^2 Z}{d r_{*}^2} + \left( \omega^2 - V_R(r) \right) Z =0 \,,
 \label{Z_radial}
\end{equation}
with the form of effective potential $V_R(r)$ is given by
\begin{equation}
    V_R(r) = N e^{-2 \sigma} \left[ 8 \pi G \left( 8 \pi G r^2 V-1  \right) \phi'^2 + 16 \pi G r \frac{d V}{d \phi} \phi' - 8 \pi G V - \frac{d^2 V}{d\phi^2} - \frac{N-1}{r^2} \right]   \,,
 \label{V_R}
\end{equation}
where the tortoise coordinate $r_*$ is 
\begin{equation}
 \frac{d r_*}{d r} = \frac{e^{\sigma}}{N}  \,.
\end{equation}

When $\omega^2<0$, the perturbation function $Z$ is unstable which gives rise to the exponential growth with respect to time. Here solving Eq.~\eqref{Z_radial} is an eigenvalue problem with the radial mode $\omega^2$ is treated as the eigenvalue, then we obtain the mode numerically by imposing the first-order derivative of the perturbation function is zero at both boundaries, $\partial_{r}Z(r_H)=\partial_{r}Z(\infty)=0$. Additionally, an auxiliary equation $\frac{d}{d r} \omega^2=0$ is introduced which allows the imposition of the condition $Z(r_m)=1$ at some point $r_m$, typically located in between the horizon and infinity. By doing so, it enables us to obtain a nontrivial and normalizable solution for $Z$, as Eq.~\eqref{Z_radial} is homogeneous. The eigenvalue $\omega^2$ can be determined exactly when $Z$ satisfies all the asymptotic boundary conditions.

\begin{figure}
\centering
 \mbox{
 (a)
 \includegraphics[angle =0,scale=0.58]{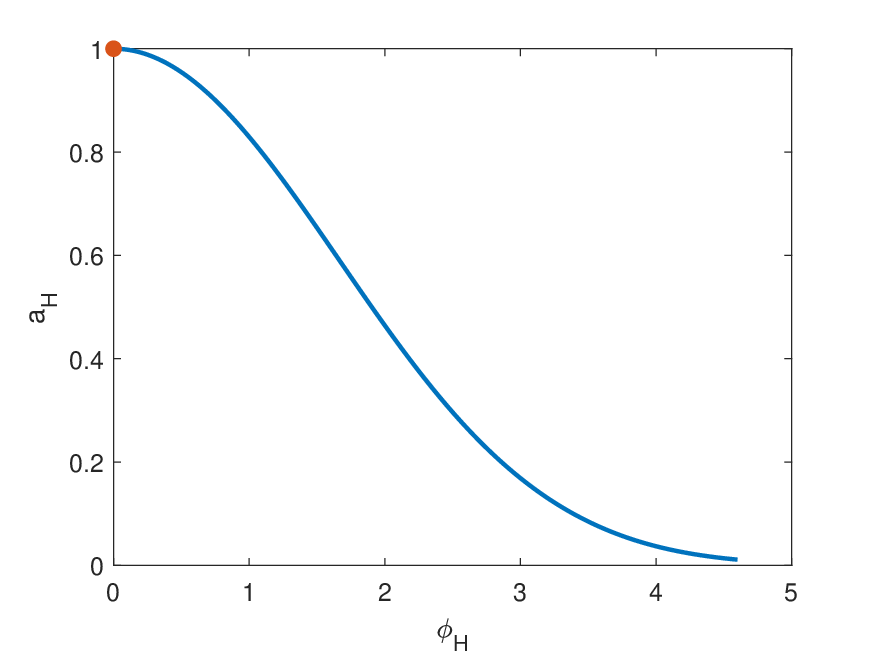}
(b)
 \includegraphics[angle =0,scale=0.58]{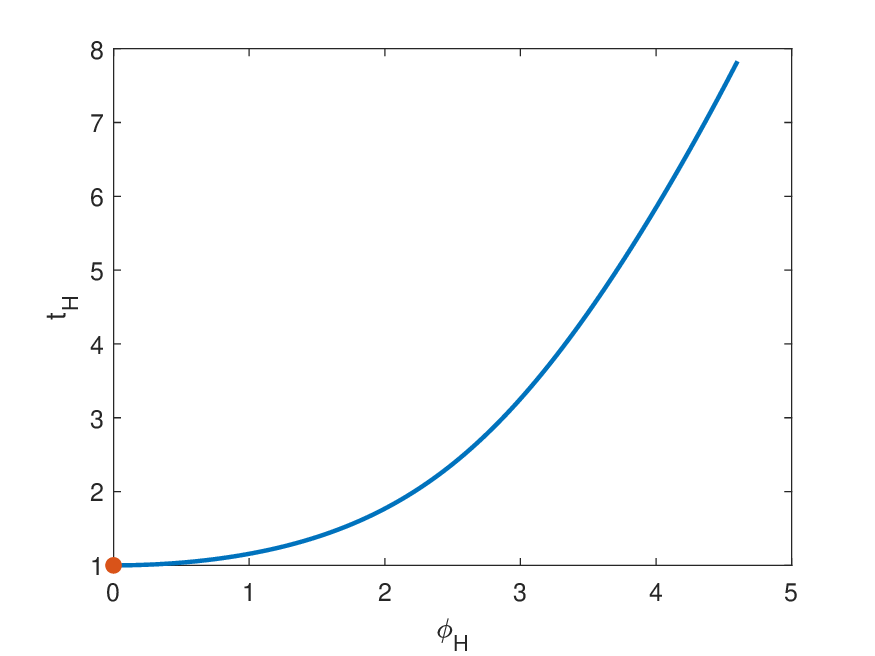}
 }
\caption{The basic properties of hairy black holes with horizon radius $r_H=1$: a) reduced area of horizon $a_H$ (b) reduced Hawking temperature $t_H$ as the function of $\phi_H$. The orange dot represents the value of Schwarzschild black hole.}
\label{plot_prop1}
\end{figure}

\section{Results and Discussions}\label{sec:res}

In the numerics, we have a few parameters which are $\Lambda$, $\mu$, $r_H$. Thus, we could introduce the following dimensionless parameters to the ODEs, 
\begin{equation}
 r \rightarrow  \frac{r}{\sqrt{\mu}} \,, \quad m \rightarrow  \frac{m}{\sqrt{\mu}} \,, \quad \phi \rightarrow \frac{\phi}{\sqrt{8 \pi G}} \,, \quad \Lambda \rightarrow 8 \pi G \Lambda \mu \,,
\end{equation}
such that we are only left with two free parameters in the calculation, which are $r_H$ and $\Lambda$. Note that the exact value of $\Lambda$ is determined when the boundary conditions have been satisfied. We also introduce the compactified coordinate $r=r_H/(1-x)$ to map the one-to-one correspondence of the horizon and infinity into $[0,1]$.

In this section, we exhibit and discuss our numerical results. Figs. \ref{plot_prop1} (a) and (b) exhibit the properties of hairy black holes with horizon radius $r_H=1$ which are the reduced area of horizon $a_H$ and reduced Hawking temperature $t_H$, respectively. Recall that $a_H=t_H=1$ (the orange dot) for the Schwarzschild black hole when the scalar field at the horizon $\phi_H$ doesn't exist. Hence, a branch of scalar hairy black holes (blue curve) emerges from the Schwarzschild black hole when $\phi_H$ exists. When $\phi_H$ starts to increase, $a_H$ decreases from 1 and then approaches zero while $t_H$ increases monotonically from 1. This indicates that the hairy black holes could possibly possess a naked singularity. These results are qualitatively similar to the case of hairy black holes supported by the asymmetric scalar potential \cite{Chew:2022enh}.

\begin{figure}
\centering
\mbox{
(a)
 \includegraphics[angle =0,scale=0.58]{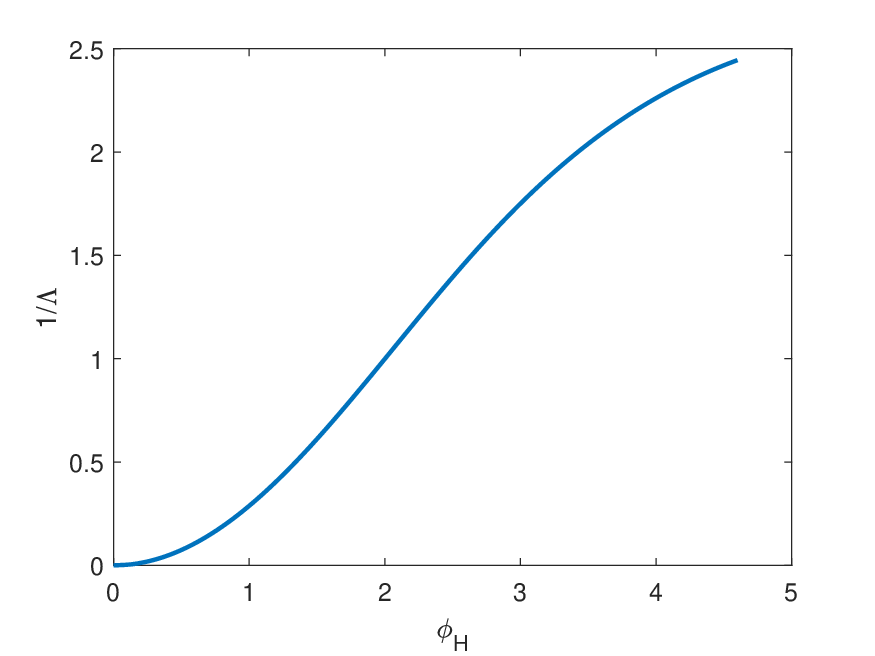}
(b)
\includegraphics[angle =0,scale=0.58]{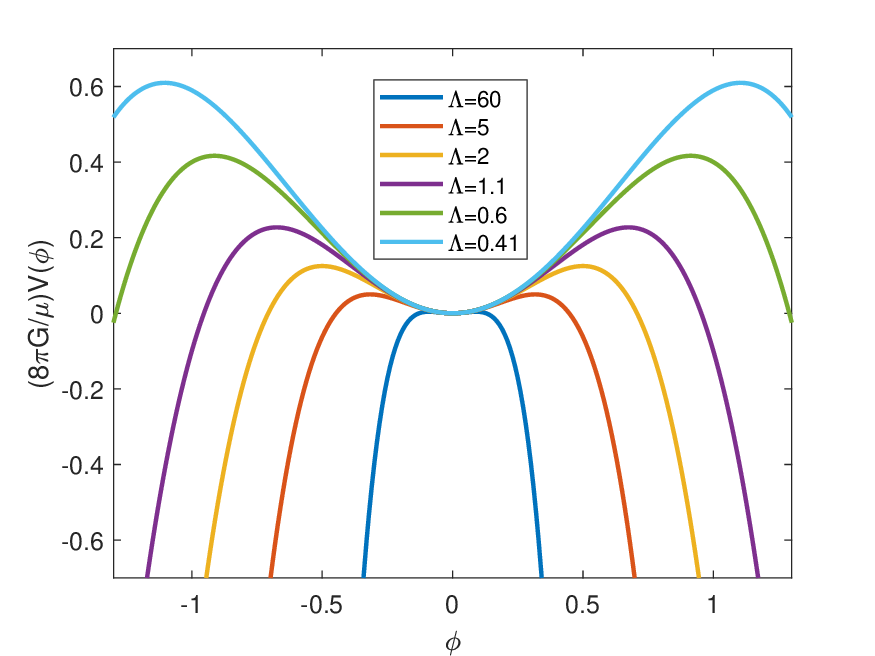}
}
\caption{(a) The relation between $\phi_H$ and $1/\Lambda$. (b) The changes of profiles of $V(\phi)$ with several values of $\Lambda$.}
\label{plot_phi_H_lambda}
\end{figure}

Fig. \ref{plot_phi_H_lambda} (a) demonstrates an inverse proportion between $\phi_H$ and $\Lambda$. As $\phi_H$ increases, $\Lambda$ decreases, and vice versa. In principle, $\Lambda$ can take any real positive values. Fig. \ref{plot_phi_H_lambda} (b) shows for small $\phi_H$ (large $\Lambda$), the height of two degenerate global maxima is very small, they are very close to each other (since $\phi_{\text{max}}=\pm \sqrt{\mu/\Lambda}$) and $\phi=0$ is the local minimum which is located in between them. Then the distance of two degenerate global maxima becomes larger and their height increase with the increasing of $\phi_H$. Note that the local minimum of $V(0)=0$ fixes the scalar field to vanish at the spatial infinity while $V(\phi_H) < 0$ and its value will become more negative (falls deeper downward along $-V$ direction) as $\phi_H$ increases.

\begin{figure}
\centering
 \mbox{
 (a)
 \includegraphics[angle =0,scale=0.58]{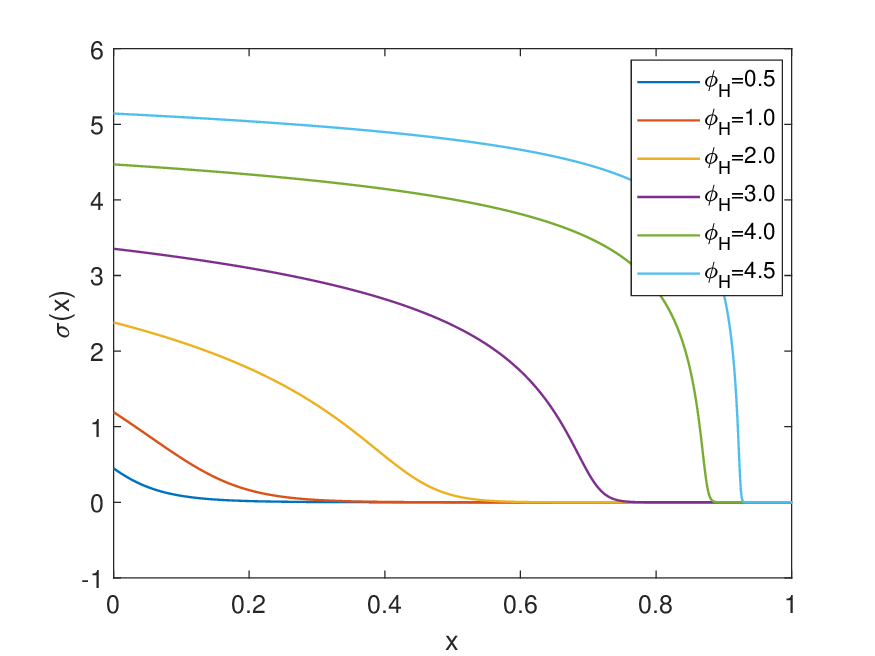}
(b)
 \includegraphics[angle =0,scale=0.58]{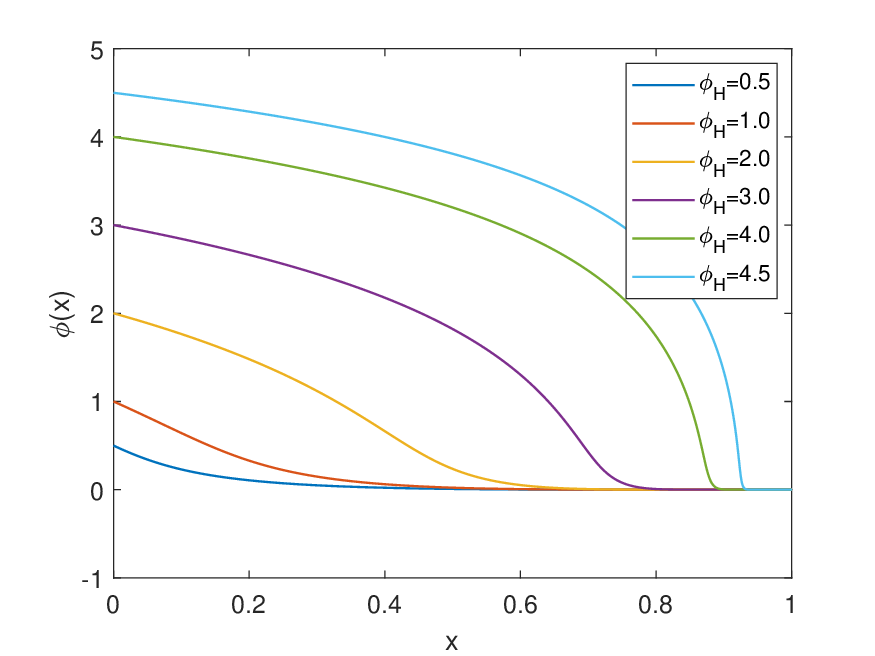}
 }
 \mbox{
(c) \includegraphics[angle =0,scale=0.58]{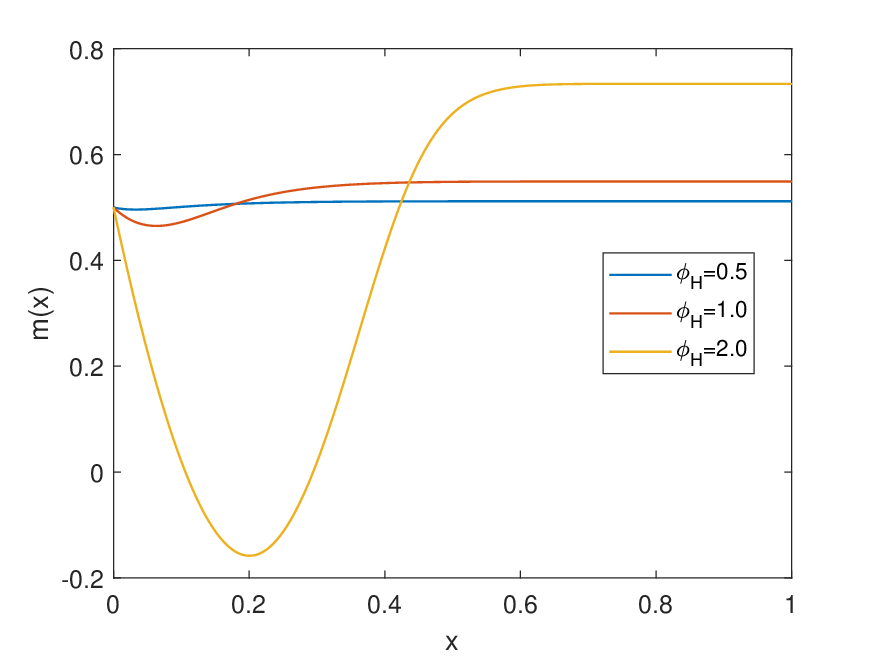}
(d) \includegraphics[angle =0,scale=0.58]{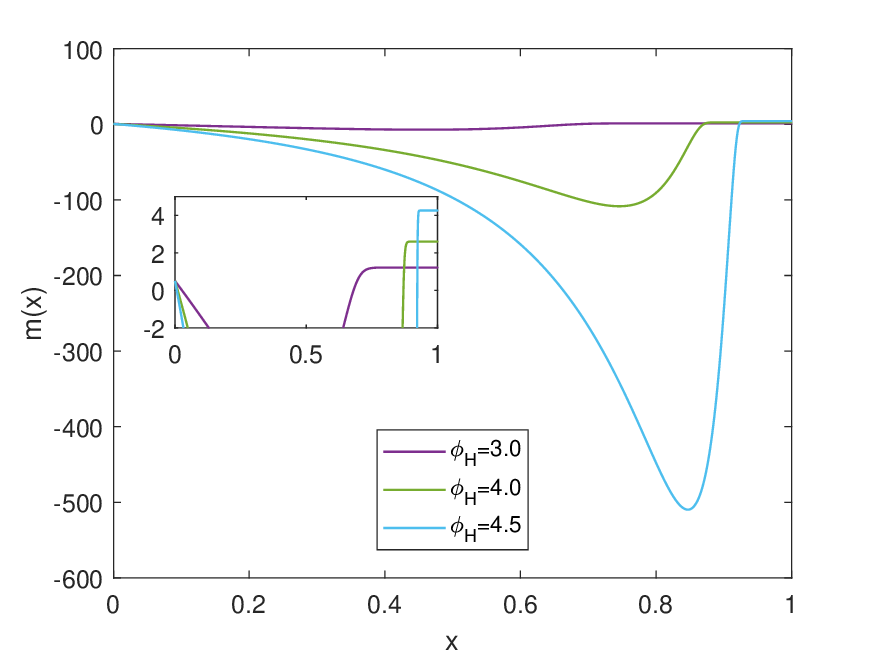}
 }
\caption{The profiles of (a) $\sigma(x)$, (b) $\phi(x)$ and (c, d) $m(x)$ solutions for the hairy black holes with $r_H=1$ and several values of $\phi_H$ in the compactified coordinate $x$.}
\label{plot_prop2}
\end{figure}

Fig. \ref{plot_prop2} (a), (b), (c) and (d) show the profiles of solutions of the hairy black holes with $r_H=1$ and several values of $\phi_H$ for the functions $\sigma(x)$, $\phi(x)$ and $m(x)$, respectively in the compactified coordinate $x$. The functions $\sigma(x)$ and $\phi(x)$ behave monotonically decreasing from their maximum value at the horizon to zero at the infinity. However, the mass function $m(x)$ decreases to a global minimum and then increases again to a constant value which is the ADM mass at spatial infinity. Note that the global minimum of mass function can become negative and be decreased to a very small number when $\phi_H$ is very large, this indicates the violation of energy condition, which will be discussed in the next paragraph. Overall, the profiles of hairy black hole solutions also behave qualitatively similar to the case of hairy black holes with asymmetric scalar potential \cite{Chew:2022enh}. 

\begin{figure}
\centering
 \mbox{
 (a)
 \includegraphics[angle =0,scale=0.58]{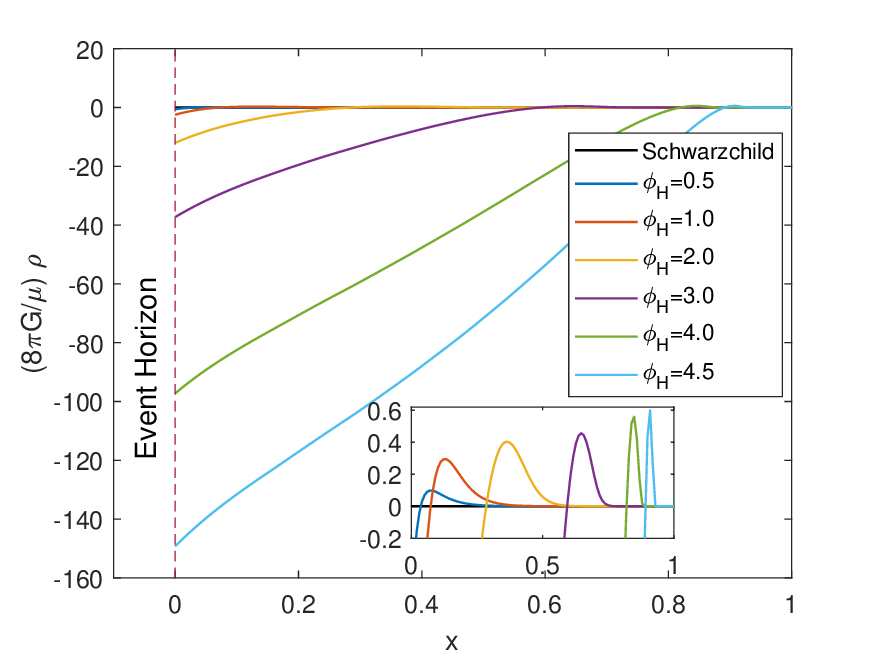}
(b)
 \includegraphics[angle =0,scale=0.58]{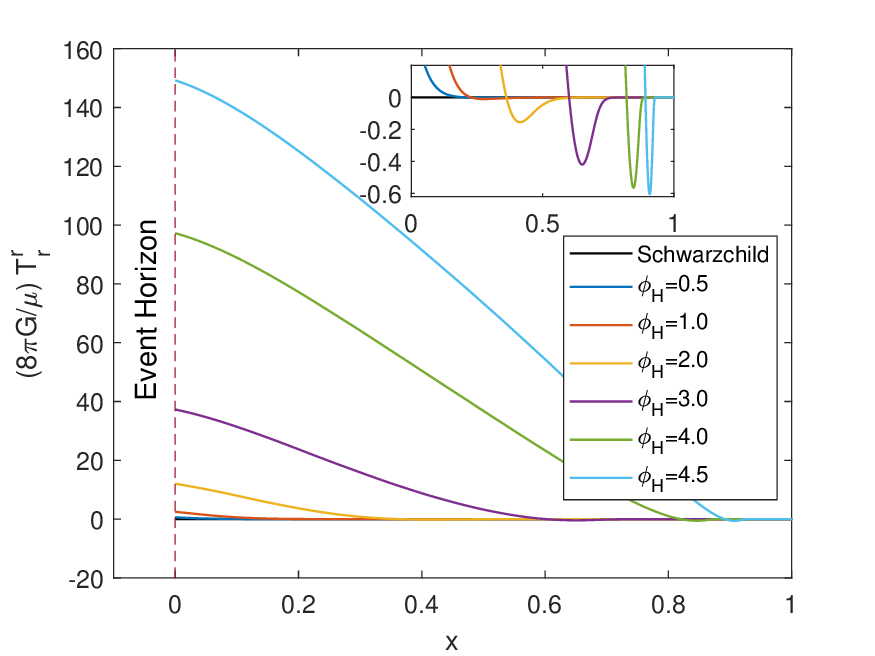}
 }
\caption{The scaled energy conditions of (a) $\frac{8 \pi G}{\mu}\rho$ and (b) $\frac{8 \pi G}{\mu}T^r_r$ for the hairy black holes with $r_H=1$ and several values of $\phi_H$.}
\label{plot_energy_cond}
\end{figure}

Fig. \ref{plot_energy_cond} (a) shows the weak energy condition (WEC) of Eq. \eqref{WEC} with several values of $\phi_H$ in the compactified coordinate $x$. We observe that the WEC is violated due to the local energy density being negative $(\rho=T^t_t<0)$ and the violation increases with the increasing of $\phi_H$, particularly at the horizon. However, the inset shows there is a small positive peak which is located exactly at the global minimum of $m(x)$ and the peak is shifted to approach the spatial infinity with the increasing of $\phi_H$. Note that the height of the peak increases with the increasing of $\phi_H$. Meanwhile, the energy condition of $T^r_r$ is slightly violated where $T^r_r$ possesses a global minimum which is negative, although large portion of $T^r_r$ is positive and greater than zero. The location of the global minimum of $T^r_r$ is also exactly the same as the location of the local minimum of $m(x)$. Similarly, the global minimum will be shifted toward the spatial infinity and decreased with the increasing of $\phi_H$.

\begin{figure}
\centering
\mbox{
 (a)
\includegraphics[angle =0,scale=0.58]{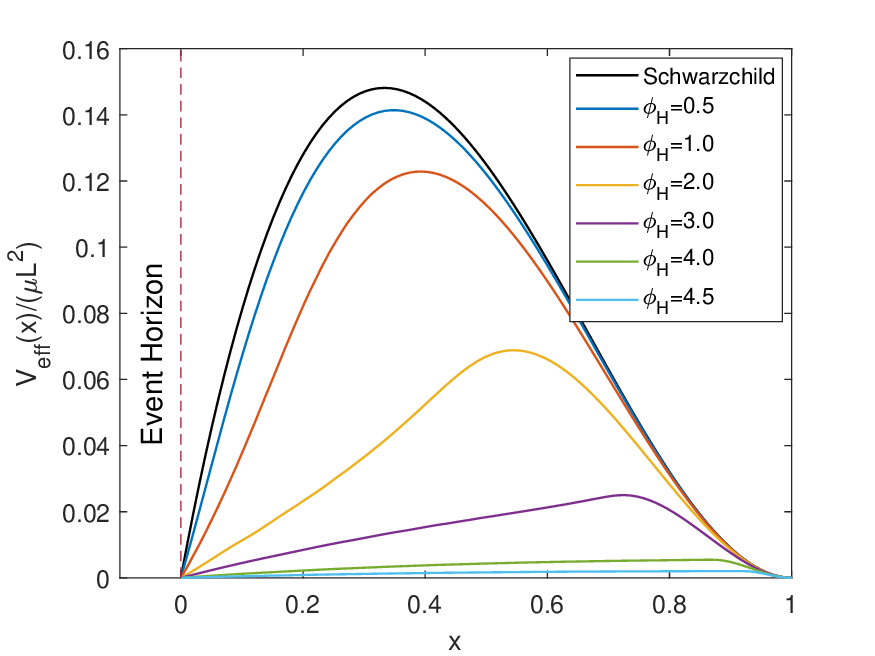}
(b)
\includegraphics[angle =0,scale=0.58]{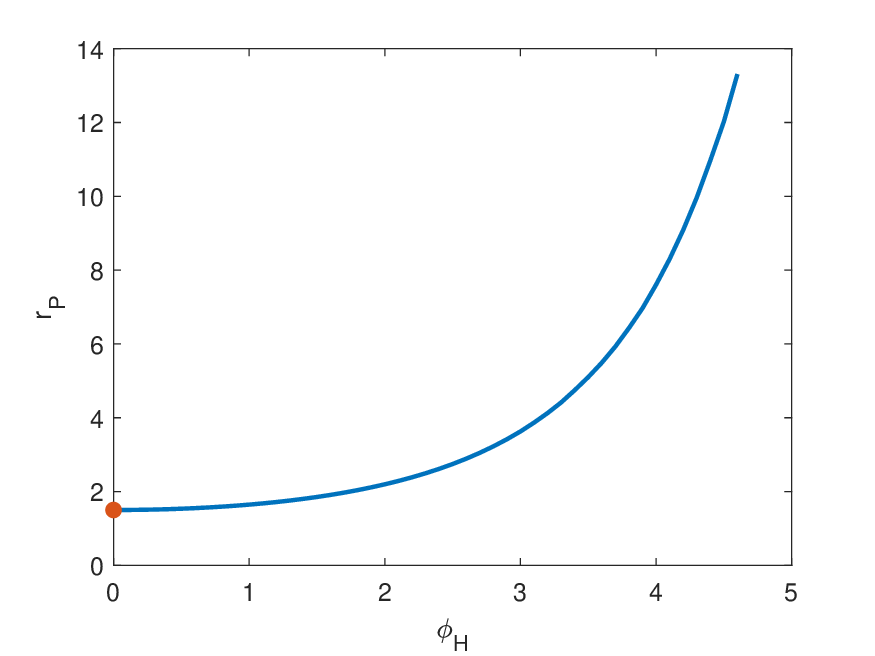}
 }
\caption{(a) The scaled effective potential of the light ring $V_{\text{eff}}(x)/(\mu L^2)$ with several values of $\phi_H$ in the compactified coordinate $x$ and the corresponding (b) location of photon sphere $r_{\mathrm{P}}$ as the function of the scalar field at horizon $\phi_H$ for the hairy black holes with $r_H=1$. The orange dot represents the value of Schwarzschild black hole.}
\label{plot_prop4}
\end{figure}

Fig. \ref{plot_prop4} (a) exhibits the scaled effective potential for photon sphere $V_{\text{eff}}(x)/L^2$ in the spacetime of hairy black holes with $r_H=1$ and several values of $\phi_H$ in the compactified coordinate $x$. $V_{\text{eff}}(x)/L^2$ possesses a peak, which indicates the existence of an unstable light ring. When $\phi_H=0$, $V_{\text{eff}}(x)/L^2$ for Schwarzschild black hole (black curve) is located at $r_{p}=1.5 r_H$ (or $x_{p}=1/3$ in the compactified coordinate), we observe that its peak is the highest. When $\phi_H$ increases, the peak of $V_{\text{eff}}/L^2$ decreases and is shifted to approach to spatial infinity. Furthermore, Fig. \ref{plot_prop4} (b) shows the location of photon sphere $r_p$ in the hairy black holes that deviate from the Schwarzschild black hole (orange dot) and increases monotonically as $\phi_H$ increases.

\begin{figure}
\centering
\mbox{
 (a)
\includegraphics[angle =0,scale=0.58]{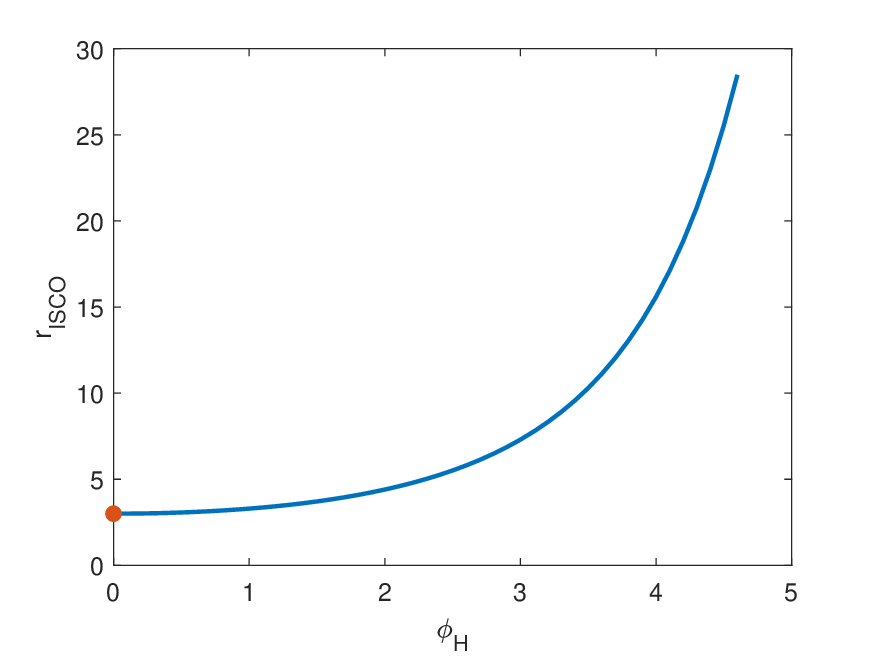}
(b)
 \includegraphics[angle =0,scale=0.58]{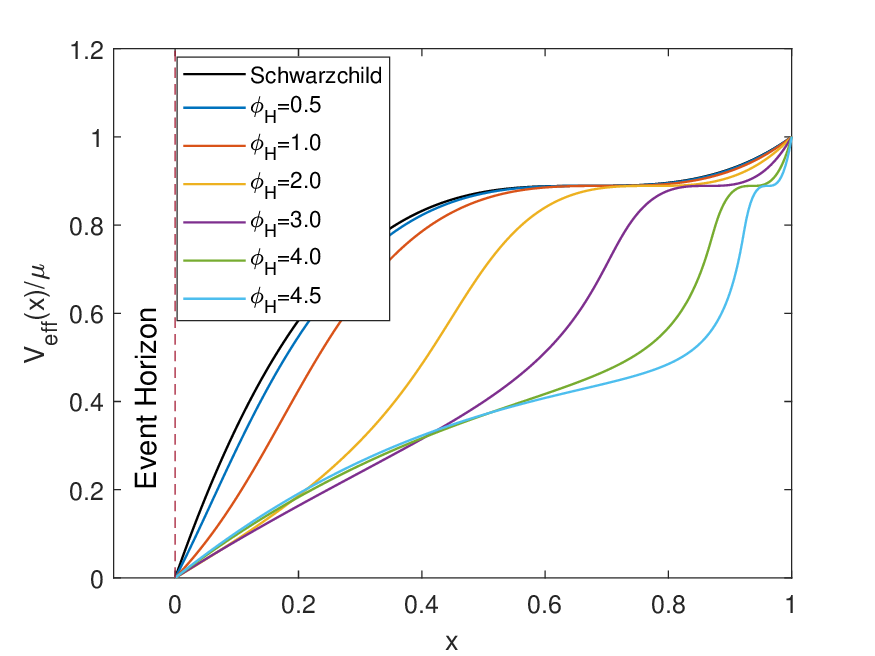}
 }
\caption{(a) The location of the innermost stable circular orbit (ISCO) of a massive test particle, $r_{\mathrm{ISCO}}$ with the corresponding (b) scaled effective potential $V_{\text{eff}}(x)/\mu$ in the compactified coordinate $x$ for the hairy black holes with $r_H=1$ and several values of $\phi_H$. The orange dot represents the value of Schwarzschild black hole.} 
\label{plot_prop3}
\end{figure}

Figs. \ref{plot_prop3} (a) shows the location of the innermost stable circular orbit (ISCO) for a massive test particle in the hairy black holes with $r_H=1$. When $\phi_H=0$, the location of ISCO for the Schwarzschild black hole is $r_{\text{ISCO}}=3 r_H$ (the orange dot). When $\phi_H$ increases from zero, the location of ISCO starts to increase and deviate from the Schwarzschild black hole. Meanwhile, Fig. \ref{plot_prop3} (b) exhibits the effective potential for a massive test particle $V_{\text{eff}}(x)/\mu$ in the hairy black holes with $r_H=1$ and several values of $\phi_H$ in the compactified coordinate $x$, we see that there is a inflection point $x_{\text{ISCO}}$ that causes $V'_{\text{eff}}(x)=V''_{\text{eff}}(x)=0$ where $V_{\text{eff}}(x)/\mu$ changes its concavity.

\begin{figure}
\centering
\mbox{
 (a)
\includegraphics[angle =0,scale=0.58]{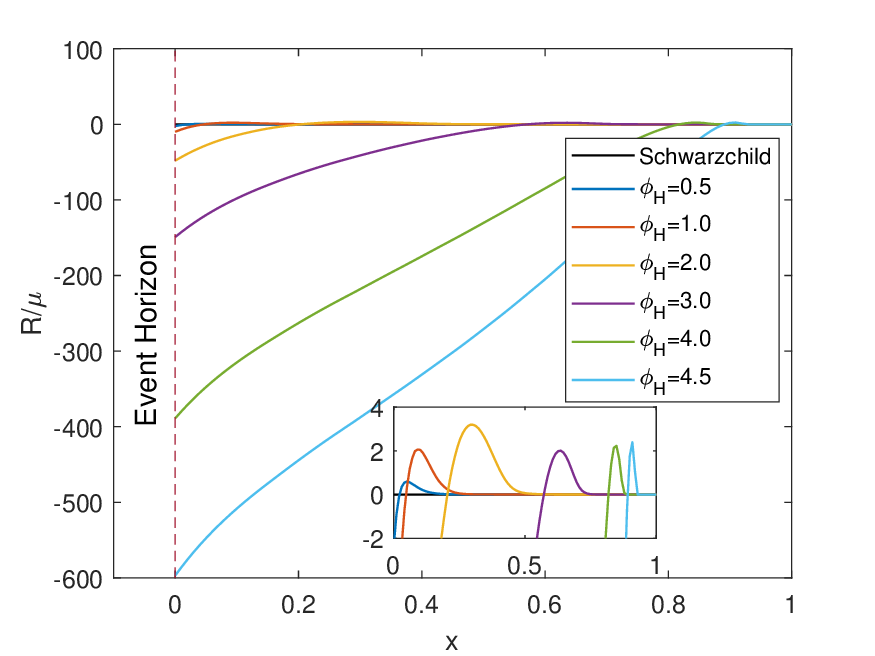}
(b)
 \includegraphics[angle =0,scale=0.58]{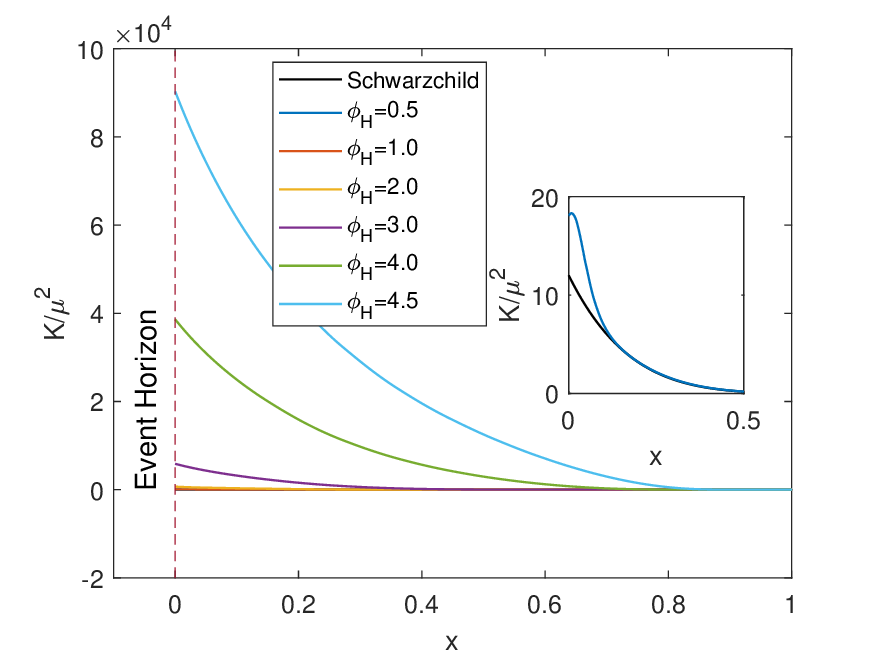}
 }
\caption{(a) The scaled Ricci scalar $R/\mu$ and (b) The scaled Kretschmann scalar $K/\mu^2$ in the compactified coordinate $x$ for the hairy black holes with $r_H=1$ and several values of $\phi_H$.}
\label{plot_ricci}
\end{figure}

Figs. \ref{plot_ricci}~(a) and (b) show the Ricci scalar $R$ and Kretschmann scalar $K$ for the hairy black holes with $r_H=1$, respectively with several values of $\phi_H$. Recall that $R=0$ and $K=12r_H^2/r^6$ for the Schwarschild black hole. When $\phi_H$ increases, the Ricci scalar $R$ of hairy black holes is non-vanishing but is negative, due to the presence of scalar hair. The inset of Fig. \ref{plot_ricci}~(a) shows a small positive peak of $R$ which is located exactly at the global minimum of $m(x)$ and the peak is shifted to the approach the spatial infinity with the increasing of $\phi_H$. Meanwhile, the Kretschmann scalar $K$ for the hairy black holes decreases monotonically to zero from its maximum value at the horizon. The maximum value of $K$ increases very sharply with the increase of $\phi_H$, which indicates the formation of naked singularity at the horizon.

\begin{figure}
\centering
\mbox{
 (a)
\includegraphics[angle =0,scale=0.58]{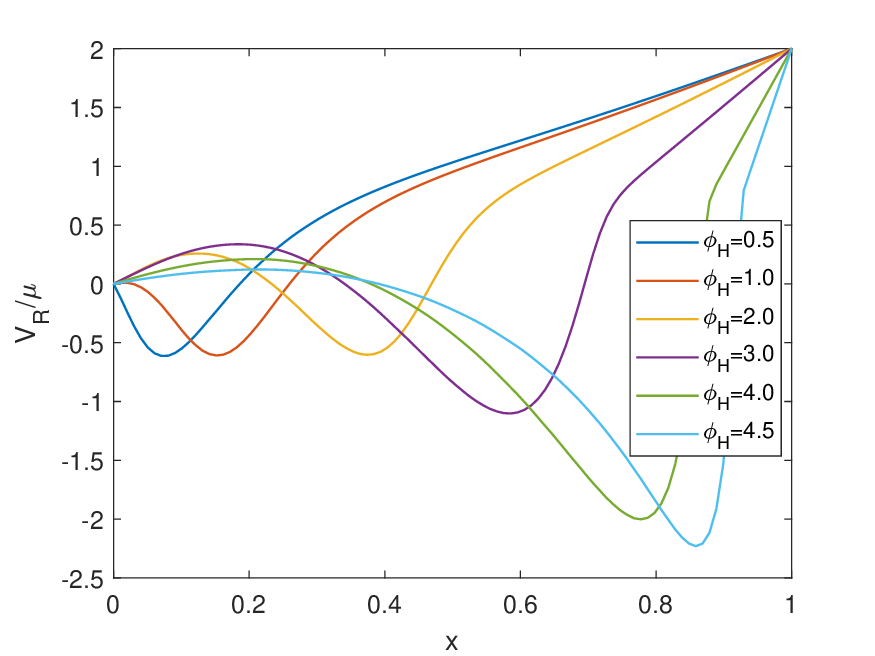}
(b)
 \includegraphics[angle =0,scale=0.58]{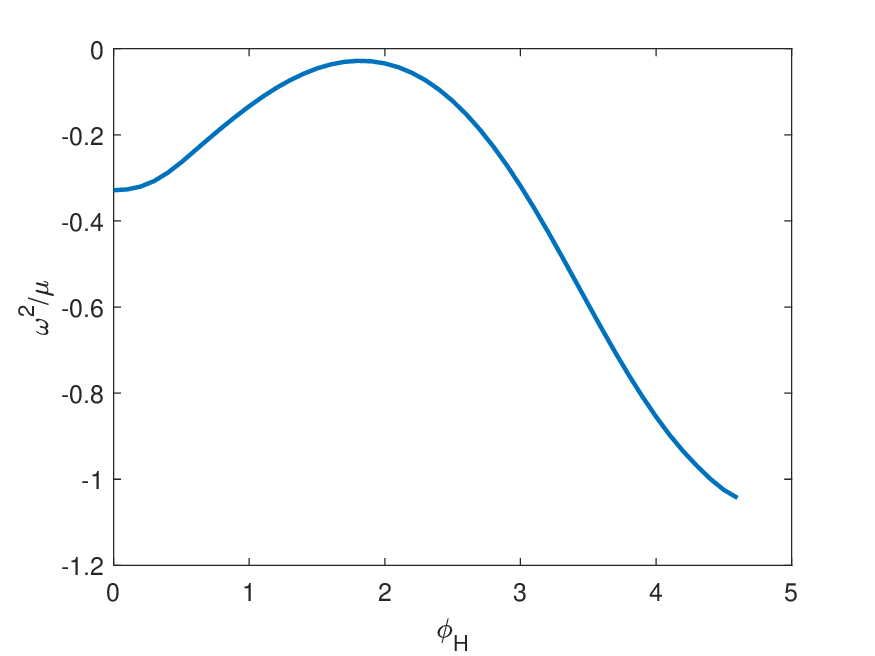}
 }
\caption{(a) The scaled effective potential $V_R(x)/\mu$ with several values of $\phi_H$ in the compactified coordinate $x$ for the radial perturbation. (b) The scaled unstable modes $\omega^2/\mu$ as the function of $\phi_H$.}
\label{plot_prop5}
\end{figure}

Fig. \ref{plot_prop5} (a) exhibits the effective potential $V_R/\mu$ in the Schr\"odinger-like master equation (Eq. \eqref{V_R}) with several $\phi_H$ in the compactified coordinate $x$. We observe that $V_R$ is negative in some regions of compactified coordinate $x$ which indicates the possibility of hairy black holes having the unstable modes. Then, we numerically perform a mode analysis to show that the hairy black holes possess the unstable radial modes with $\omega^2<0$ as shown in Fig. \ref{plot_prop5} (b) where $\omega^2$ increases to maximum value then decreases again. Hence, the configuration of hairy black holes is unstable since the perturbation grows exponentially with time. Moreover, their counterpart hairy black holes supported by an asymmetric scalar potential are also unstable against the radial perturbation \cite{Chew:2022enh}.

\section{Conclusion}\label{sec:con}

We numerically construct the asymptotically flat solutions of hairy black holes by minimally coupling the Einstein gravity with a scalar potential which contains a quartic self-interaction term. The potential looks like an inverted Mexican hat with two degenerate global maxima and a local minimum is located in between them. Here we fix the local minimum as zero to ensure the scalar field vanishes at the spatial infinity. The trivial solution is the Schwarzschild black hole when we switch off the scalar field at the horizon, $r_H$. When we increase the strength of the scalar field at the horizon $\phi_H$, a branch of hairy black hole solutions emerges from the Schwarzschild black hole. 

In this paper, we perform a comprehensive study on the properties of hairy black holes. First, we study the area of horizon $A_H$ and Hawking temperature $T_H$ of hairy black holes, hence we introduce the reduced area of horizon $a_H$ and reduced Hawking temperature  $t_H$ to rescale $A_H$ and $T_H$, respectively, in such a way that we can compare our hairy black holes solutions with a known solution, which is Schwarzschild black hole in this case. Recall that $a_H=t_H=1$ for Schwarzschild black hole. When we increase $\phi_H$, $a_H$ decreases from 1 to zero while $t_H$ increases sharply from 1, this implies that the hairy black holes could possibly possess a naked singularity at the horizon. On the other hand, the coefficient for the quartic self-interaction term in the scalar potential can take any real positive values and is inversely proportional to $\phi_H$. Initially the two degenerate global maxima of scalar potential are very close to each other and their height are very small for small $\phi_H$. However, their separation become wider and height increase with the increasing of $\phi_H$.

Secondly, the profiles of all solutions except the mass function for the hairy black holes behave monotonically, they decrease monotonically from a maximal value at the horizon to zero at the spatial infinity. However, the mass function decreases from its Schwarzschild mass value at the horizon to a global minimum value and then increases very sharply to reach a constant value which is the ADM mass at the spatial infinity. When $\phi_H$ increases, the global minimum of mass function can possess some negative values and can be further decreased with the increasing of $\phi_H$, this indicates the violation of weak energy condition. Moreover, the profiles of solutions are qualitatively similar to the hairy black holes supported by an asymmetric scalar potential \cite{Chew:2022enh}.

Thirdly, we study the weak energy condition (WEC) of hairy black holes. The WEC of hairy black holes also represents the local energy density $\rho$ but it is violated, particularly it is most severely violated at the horizon, and the violation increases with the increasing of $\phi_H$. However, $\rho$ possesses a peak which is positive and located exactly at the location of global minimum of mass function. The peak will be shifted toward the spatial infinity with the increasing of $\phi_H$. Although the energy condition for the component of stress-energy tensor $T^r_r$ is positive in the large portion of radial coordinate $r$, it is slightly violated since its global minimum becomes negative and has been developed exactly at the location of global minimum of mass function.

We also study the geodesic of test particles in the vicinity of hairy black holes by calculating the innermost stable circular orbits (ISCO) for massive test particle and location of photon sphere. Recall that the location of ISCO is at $3r_H$ and the location of the photon sphere is at $1.5r_H$ for the Schwarzschild black hole. When $\phi_H$ increases, the location of ISCO and photon sphere start to deviate and increase monotonically from the Schwarzschild black hole.

Moreover, we study the Ricci scalar and Kretschmann scalar of hairy black holes. The Schwarzschild black hole possesses the vanishing Ricci scalar and the Kretschmann scalar with the expression $12r^2_H/r^6$. When $\phi_H$ increases, the Ricci scalar of hairy black holes is negative in almost the entire region but has a peak which is positive and located at the location of global minimum of the mass function while the Kretschmann scalar is entirely positive and decreases monotonically from the horizon to zero at the spatial infinity. The Kretschmann scalar at the horizon increases with the increasing of $\phi_H$, which indicates the formation of naked singularity at the horizon.

Lastly, we perform a linear perturbation to the hairy black holes' spacetime and scalar field to derive a Schr\"odinger-like master equation to study the linear stability of such configuration. The linear stability of hairy black holes can be determined numerically by the mode analysis where the hairy black holes are found to be unstable against the linear perturbation. Their counterpart hairy black holes with an asymmetric scalar potential are also unstable against the linear perturbation \cite{Chew:2022enh}.

\section*{Acknowledgement}
XYC acknowledges the support from the starting grant of Jiangsu University of Science and Technology. XYC is also grateful for the hospitality from Dong-han Yeom during the visit to Pusan National University and from Gansukh Tumurtushaa during the visit to Jeju National University.

\end{document}